\documentclass{emulateapj}
\usepackage{apjfonts}

\newcommand{\Msun}      {\mbox{$\rm\,M_{\mathord\odot}$}}

\submitted{Accepted by the Astrophysical Journal}

\begin{document}

\lefthead{XTE J1650--500 at Low Luminosity}
\righthead{J.A. Tomsick et al.}

\title{Detection of Low-Hard State Spectral and Timing 
Signatures from the Black Hole X-Ray Transient XTE J1650--500 
at Low X-Ray Luminosities}

\author{John A. Tomsick\altaffilmark{1},
Emrah Kalemci\altaffilmark{2},
Philip Kaaret\altaffilmark{3}}

\altaffiltext{1}{Center for Astrophysics and Space Sciences, Code
0424, University of California at San Diego, La Jolla, CA,
92093, USA (e-mail: jtomsick@ucsd.edu)}

\altaffiltext{2}{Space Sciences Laboratory, University of 
California, Berkeley, CA, 94720-7450, USA}

\altaffiltext{3}{Harvard-Smithsonian Center for Astrophysics, 
60 Garden Street, Cambridge, MA, 02138, USA}

\begin{abstract}

Using the {\em Chandra X-ray Observatory} and the {\em Rossi X-ray 
Timing Explorer}, we have studied the black hole candidate (BHC) 
X-ray transient XTE J1650--500 near the end of its 2001-2002 outburst 
after its transition to the low-hard state at X-ray luminosities down 
to $L = 1.5\times 10^{34}$ erg~s$^{-1}$ (1-9 keV, assuming a source 
distance of 4 kpc).  Our results include a characterization of the 
spectral and timing properties.  At the lowest sampled luminosity, we 
used an 18 ks {\em Chandra} observation to measure the power spectrum 
at low frequencies.  For the 3 epochs at which we obtained
{\em Chandra}/{\em RXTE} observations, the 0.5-20~keV energy spectrum 
is consistent with a spectral model consisting of a power-law with 
interstellar absorption.  We detect evolution in the power-law photon 
index from $\Gamma = 1.66\pm 0.05$ to $\Gamma = 1.93\pm 0.13$ (90\% 
confidence errors), indicating that the source softens at low 
luminosities.  The power spectra are characterized by strong (20-35\% 
fractional rms) band-limited noise, which we model as a zero-centered 
Lorentzian.  Including results from an {\em RXTE} study of XTE 
J1650--500 near the transition to the low-hard state by Kalemci et al. 
(2003), the half-width of the zero-centered Lorentzian (roughly where 
the band-limited noise cuts off) drops from 4 Hz at 
$L = 7\times 10^{36}$ erg~s$^{-1}$ (1-9 keV, absorbed) to $0.067\pm 
0.007$ Hz at $L = 9\times 10^{34}$ erg~s$^{-1}$ to $0.0035\pm 0.0010$ 
Hz at the lowest luminosity.  While the spectral and timing parameters 
evolve with luminosity, it is notable that the general shapes of the 
energy and power spectra remain the same, indicating that the source 
stays in the low-hard state.  This implies that the X-ray emitting 
region of the system likely keeps the same overall structure, while 
the luminosity changes by a factor of 470.  We discuss how these
results may constrain theoretical black hole accretion models.

\end{abstract}


\section{Introduction}

To fully understand the behavior of black hole candidate (BHC) 
X-ray transients, it is important to determine the source
properties for the full range of observed X-ray luminosities
from outburst peak at $L\sim 10^{38-39}$ erg~s$^{-1}$ to 
quiescence at $L\sim 10^{30-33}$ erg~s$^{-1}$.  The brighter 
part of this range has been well-studied by several satellites, 
and, due to its broad bandpass (2-200 keV) and its excellent 
timing capabilities, the {\em Rossi X-ray Timing Explorer} 
({\em RXTE}, Bradt, Rothschild \& Swank 1993\nocite{brs93})
has proved to be an extremely useful tool for studying
the luminosity range above $\sim$$10^{35-36}$ erg~s$^{-1}$
(for sources with typical distances of 3-10~kpc).  For most 
systems, this is below the luminosity where the source makes 
a transition from the canonical high-soft state to the 
low-hard (or hard) state \citep{mr03}, and our group as well 
as others have used {\em RXTE} to study BHC transients as 
they undergo soft-to-hard state transitions during outburst 
decay \citep{tk00,nwd02,kalemci_thesis,kalemci03a}.  However, 
as {\em RXTE} is not an imaging instrument, background and 
source confusion become significant at lower luminosities, 
making both spectral and timing studies problematic.  Thus, 
at low luminosities, imaging observatories such as {\em BeppoSAX}, 
the {\em Chandra X-ray Observatory} \citep{weisskopf02}, and 
{\em XMM-Newton} are necessary, and these satellites have been 
used to study BHC transients in quiescence \citep{garcia01}.  
The aim of this study is to use {\em Chandra} to study the 
evolution of XTE J1650--500 in the luminosity range between 
the transition to the hard state and quiescence.

Few previous studies have been dedicated to observing BHC 
transients at luminosities between the hard state transition 
and quiescence even though the source evolution has not been 
well-characterized, and the physical changes that occur in 
this regime are unclear.  The results of the studies of BHC 
transients that have been undertaken in this regime are 
intriguing.  A correlation between X-ray and radio flux has 
been found for several BHC systems in the hard state 
\citep{corbel03,gfp03}, and this has been interpreted as 
evidence for a connection between X-ray production and a 
compact radio jet \citep{markoff03}.  As the sources decay, 
it has been observed that the X-ray energy spectra maintain 
a power-law shape; however, in some cases, there is evidence
for spectral evolution with sources becoming softer at low 
luminosities \citep{ebisawa94,tck01,corbel03,kalemci_thesis}.
X-ray timing studies immediately after the hard state 
transition show that, in most cases, the characteristic
frequencies drop \citep{tk00,nwd02,kalemci03a}, and this may 
indicate movement of the inner edge of the optically thick 
accretion disk away from the black hole.  However, for the 
most part, the X-ray timing studies have not extended to 
luminosities much below the hard state transition.

The hard state accretion geometry, the emission mechanisms, 
and the connection between the accreting material and the
compact jet outflow that is present in the hard state for 
at least some BHCs \citep{fender01} are all currently
active areas of debate.  Several theoretical models have
been suggested.  One possibility is that a quasi-spherical, 
optically thin region forms in the inner portion of the 
accretion flow as in advection-dominated accretion flow 
(ADAF) models \citep{nmy96}.  The X-ray emission is
predominantly due to thermal Comptonization in the hard 
state for ADAF models and for more generic ``sphere+disk'' 
models, and some success has been achieved in comparing
such models to observations \citep{emn97,nowak99}.  
However, for a range of disk viscosities, the ADAF model 
is convectively unstable \citep{qg00}, and this should be 
considered before concluding that any ADAF or sphere+disk 
model provides the correct physical description.  
Another possibility that implies a much different site 
for X-ray production and a different accretion geometry 
is that the X-ray emission is due to magnetic flares 
above the disk \citep{grv79,dcf99}.  Although the 
presence of a jet is not incorporated into the basic 
models described above, some recent work describes how 
a jet may arise from a ``magnetic corona'' \citep{mf02}.  
Other models have focused on the emission from the 
compact jet, and one suggestion is that the X-ray 
emission is produced in the jet via a synchrotron 
mechanism \citep{mff01}.  Improving the luminosity coverage 
for the hard state is important for providing tests of these 
models.  In addition, it is currently unclear whether the 
same physical model can apply to these systems in the hard
state and in quiescence.  Major physical changes to the 
system could be seen as another state transition at very 
low luminosities.  On the other hand, the lack of a sharp
transition could indicate that quiescence is simply a
low luminosity extension of the hard state as has been
previously suggested for GX 339--4 \citep{corbel00,kong00}.

We chose to carry out our program by observing XTE J1650--500.  
This was a new BHC transient when it was discovered in 
outburst in 2001 \citep{remillard01}.  Although the compact 
object mass is not well-constrained so that it has not been 
confirmed that this is a black hole system, its X-ray 
properties strongly suggest that it harbors a black hole 
\citep{tomsick03,homan03,kalemci03b}.  During its outburst, 
the source was optically identified \citep{ct01}, and a radio 
counterpart was found \citep{groot01}.  In quiescence, a radial 
velocity study was carried out, indicating a binary orbital 
period of 5.1~hr and an optical mass function of 
$0.64\pm 0.03$~\Msun~\citep{sf02}.  The value for the mass 
function is relatively low for a black hole system and could 
indicate that the system contains a relatively low mass black 
hole, has a low binary inclination ($i<40^{\circ}$), or both.
We previously reported on {\em RXTE} observations of 
XTE J1650--500, including a study of its timing properties near 
the soft-to-hard state transition \citep{kalemci03b}.  We also 
reported on 14 day X-ray oscillations and $\sim$100~s X-ray 
flares that were detected with {\em RXTE} near the end of the 
outburst \citep{tomsick03}.  The current work focuses on our 
studies of the source with {\em Chandra}.

\section{Observations and Analysis}

\begin{figure}
\centerline{\includegraphics[width=0.5\textwidth]{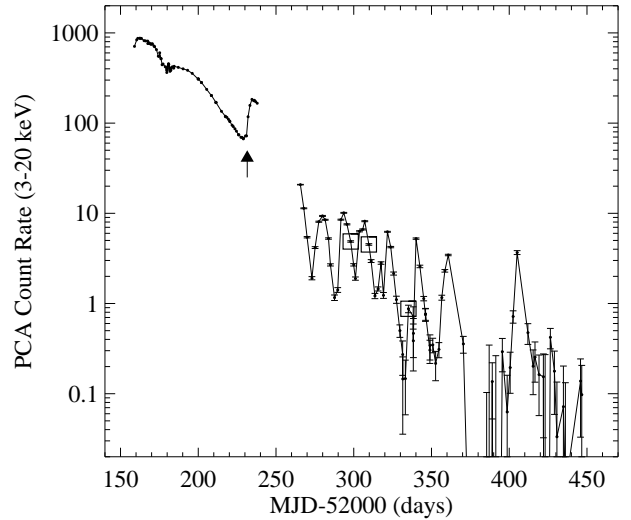}}
\vspace{0.2cm}
\caption{The 3-20 keV light curve for the 2001-2002 XTE J1650--500
outburst using PCA data from all the {\em RXTE} observations.  The
time of the soft-to-hard state transition, which occurred on MJD 
52231.5, is marked with an arrow, and the gap in observations that 
is centered near MJD 52250 is due to solar angle constraints.  After 
the gap, the source exhibited 14 day oscillations \citep{tomsick03}.  
The 3 {\em Chandra} observations occurred during these oscillations
and are marked with squares.\label{fig:lc}}
\end{figure}

We obtained 3 {\em Chandra} observations of XTE J1650--500 
near the end of its 2001-2002 outburst, and the times of the 
observations are marked on the long-term X-ray light curve 
shown in Figure~\ref{fig:lc}.  We compiled the 3-20 keV light 
curve in Figure~\ref{fig:lc} using all the observations of 
XTE J1650--500 that were made with the Proportional Counter 
Array (PCA) on {\em RXTE}.  The 3 {\em Chandra} observations 
were made 67, 78, and 104 days after the soft-to-hard state 
transition \citep{homan03,kalemci03b}, during the time when 
the source was exhibiting large-amplitude flux oscillations 
with a characteristic time scale of 14 days \citep{tomsick03}.  
All 3 observations were made at flux levels well below the 
level of the soft-to-hard state transition.  

The {\em Chandra} observations were made using the Advanced CCD 
Imaging Spectrometer (ACIS).  In each case, the source was placed on 
one of the back-illuminated ACIS chips (S3).  The first two observations 
(1 and 2), were made with the High Energy Transmission Grating (HETG) 
inserted, and each observation lasted approximately 10~ks (see 
Table~\ref{tab:obs} for exact times and other detailed information
about the observations).  To increase the throughput, the grating was 
not used for observation 3, making photon pile-up \citep{davis01} a
concern.  To mitigate the effects of pile-up, we reduced the CCD
exposure time from 0.4~s per frame from the nominal 3.2~s per 
frame by using a 1/8 CCD sub-array.  In addition, the telescope was
pointed $2^{\prime}.7$ away from the target, slightly blurring the 
point spread function (PSF) for XTE J1650--500.  An exposure time
of 18~ks was obtained for observation 3.  In addition to the 
{\em Chandra} observations, shorter 1.8-2.4~ks {\em RXTE} observations
were co-ordinated to cover part of each {\em Chandra} observation.

\subsection{{\em Chandra} Analysis}

For the two grating observations, we used the ``level 2'' event 
list produced by the standard data processing (ASCDS v6.5.1 using 
CALDB v2.11).  For both observations, we applied a standard 
correction to the ACIS-S4 chip to correct a flaw in the chip 
readout.  For this, as well as the data processing described below, 
we used the CIAO ({\em Chandra Interactive Analysis of 
Observations}) v2.3 software and CALDB (Calibration Data Base) 
v2.21.  Prior to extracting the grating spectra, we produced 
background light curves.  For observation 1 only, we found a 
few background flares where the count rate increased by 50-100\%
for a time period of less than 50~s.  Due to their short duration, 
the flares make only a small contribution to the total background.
Thus, we did not remove these times in the subsequent analysis, 
and the full integration time was used for both observations.  

For observations 1 and 2, we produced 0.3-8 keV zero order
images.  We used the source detection algorithm {\em wavdetect} 
to search for sources in a 2 by 2 arcminute region around the 
{\em Chandra} aim-point, and we only detect the target in each 
case.  The observation 1 and 2 source positions are consistent 
with one another, and we obtain a mean source position of 
R.A. = 16h 50m 00s.98, 
Dec. = --49$^{\circ}$~57$^{\prime}$~43$^{\prime\prime}$.6
(equinox 2000.0) with an uncertainty of $0^{\prime\prime}.6$.  
This position is consistent with the previously reported radio 
\citep{groot01} and optical \citep{ct01} positions, and we note 
that the {\em Chandra} position is somewhat more accurate than 
the $3^{\prime\prime}$ and $2^{\prime\prime}$ radio and optical 
positions, respectively.  We conclude that the detected source 
is indeed XTE J1650--500.  Given the recent detections of X-ray 
jets from XTE J1550--564 \citep{corbel02} and 4U 1755--338 
\citep{aw03}, we examined the XTE J1650--500 images for X-ray 
extensions in the vicinity of the source, but we do not see any 
features that are likely to be jets.  Using the zero order ACIS 
response matrix and a spectral shape similar to that observed 
for the XTE J1550--564 jet ($\Gamma = 1.6$ with interstellar 
absorption), the upper limit on the 0.3-8 keV absorbed flux is 
$6\times 10^{-14}$ erg~cm$^{-2}$~s$^{-1}$ at positions 
$10^{\prime\prime}$ from XTE J1650--500.

We used the CIAO routine {\em tgextract} for the extraction
of grating spectra and the CIAO routines {\em mkgrmf} and 
{\em mkgarf} to produce response matrices.  Combining 
the +1 and $-1$ MEG (Medium Energy Grating) 0.7-7 keV spectra 
yields 4786 counts in 10001 seconds for observation 1 and 
4342 counts in 9509 seconds for observation 2.  For the 
+1 and $-1$ HEG (High Energy Grating) 1-9 keV spectra, there 
are 2706 counts and 1872 counts for observations 1 and 2, 
respectively.  We also produced estimated background spectra; 
however, we find that the background is not significant with 
estimates of $\sim$20 counts for the MEG and $\sim$11 counts 
for the HEG (0.4\% of the detected counts in both cases).  
The higher MEG and HEG orders contain between $\sim$30 and 
$\sim$200 counts (source plus background).  For each 
observation, the zeroth order contains $\sim$2500 counts, 
but there is a high level of photon pile-up.  The flux 
obtained from the grating spectra implies that the number 
of counts would be a factor of 3 higher without photon 
pile-up.  Thus, due to the low number of counts in grating 
orders 2 and 3 and the high level of pile-up in the zeroth 
order, we focus on the order 1 spectra in this work.  A 
final step in the extraction of the spectra is to apply 
a correction to the response matrix to account for the
gradually changing low energy ACIS response.  We used the
script {\em corrarf} to apply this correction.

For observation 3, we used the ``level 2'' event list 
produced by the standard data processing (ASCDS v6.6.0 and 
CALDB v2.12).  After inspecting an image including the full 
energy band, we produced a 0.3-8 keV image to reduce the 
background level.  With the 1/8 sub-array and only the S3 
chip active, the field of view is approximately $1^{\prime}$ 
by $8^{\prime}.4$ (128 by 1024 pixels), and we used 
{\em wavdetect} to search for sources.  We found 4 sources 
detected at greater than 4-$\sigma$ significance.  The 
brightest source by far is at a position consistent with 
the XTE J1650--500 position found above for the grating
observations.  Although XTE J1650--500 appears to be
elongated, we used CIAO to simulate an off-axis PSF, 
and we found that XTE J1650--500 is consistent
with a point-like source.  The other 3 sources also
appear to be point-like, and the closest source is 
more than $1^{\prime}$ from XTE J1650--500.  Despite 
using an instrument setup designed to mitigate pile-up, 
the measured ACIS count rate of 0.493~s$^{-1}$ for
XTE J1650--500 (see Table~\ref{tab:obs}) indicates
that the data at the PSF core is moderately affected 
by photon pile-up.  However, we were still able to use 
this data for spectral and timing analysis as described 
below and in Appendix A.

\subsection{{\em RXTE} Analysis}

We initially performed spectral and timing analysis for 
the 3 {\em RXTE} observations that were made during 
the {\em Chandra} observations.  Using the 
``Standard 2'' PCA data, we extracted energy spectra 
using scripts developed at UC San Diego and the 
University of T\"{u}bingen that incorporate the 
standard software for {\em RXTE} data reduction 
(FTOOLS v5.2).  We used the 2002 February release of 
the ``Faint Source'' model for background subtraction.  
Obtaining the best possible background subtraction is 
important for this work because of the low source count 
rates.  To achieve this, we used Standard 2 data from 
the top anode layer only and excluded data from PCU 0, 
which has a higher background level due to the loss of 
its propane layer\footnote{See
http://lheawww.gsfc.nasa.gov/users/craigm/pca-bkg/bkg-users.html
for a detailed analysis of the background model performance.}.

In addition to the 3 {\em RXTE} observations, we extracted 
energy spectra for the final 8 observations for which the 
PCA count rates are shown in Figure~\ref{fig:lc}.  
For these observations, the mean 5-8 keV PCA count rate 
is $0.021\pm 0.018$ s$^{-1}$ per Proportional Counter Unit 
(PCU) using all three anode layers, indicating that the
rate is consistent with zero at slightly more than the
1-$\sigma$ level.  We determined this rate in order to
make a comparison to the rate expected from the Galactic
ridge emission at the Galactic latitude of XTE J1650--500
($b = -3.44^{\circ}$).  \cite{vm98} have determined the 
level of Galactic ridge emission (in PCA count rate) 
averaged over Galactic longitude as a function of $b$
(see Figure~2 of Valinia \& Marshall 1998\nocite{vm98}).
They found that the distribution could be described by
narrow and broad gaussian components, and only the 
broad gaussian contributes at the Galactic latitude of
XTE J1650--500.  Using the parameters for the broad
gaussian leads to a best estimate of the 5-8 keV 
Galactic ridge rate of 0.71 s$^{-1}$ for all anode 
layers and 5 PCUs.  Scaling the observed count rate 
given above (0.021~s$^{-1}$) to 5 PCUs gives 
$0.11\pm 0.09$ s$^{-1}$, which is significantly below 
the best estimate for the Galactic ridge emission.  Thus, 
we conclude that the Galactic ridge emission can easily 
explain the low level of emission we detect for the final 
8 observations.  As this analysis shows that it is likely 
that XTE J1650--500 does not contribute to the flux detected 
during these observations, we produced an average spectrum 
(top anode layer only) for the final 8 observations and 
subtracted it from the PCA spectra for the 3 {\em RXTE} 
observations that were simultaneous with the {\em Chandra} 
observations.  The effect to the continuum is small; 
however, the Galactic ridge emission contains a strong 
iron emission line at 6.7 keV, and this line is apparent 
in the average spectrum for the final 8 observations.

Finally, we analyzed the data from the 15 {\em RXTE} 
observations taken between MJD 52265 and MJD 52298, which is 
after the gap in {\em RXTE} coverage and before the 
{\em Chandra} observations began (see Figure~\ref{fig:lc}).  
We processed the PCA data as described above, including the 
subtraction of the Galactic ridge emission from the energy 
spectrum.  For the spectral analysis described in \S~3.3, 
we also removed flares from 2 of the observations as 
described in \cite{tomsick03}.

\section{Results}

\subsection{Energy Spectra at Low Flux Levels}

\begin{figure}
\centerline{\includegraphics[width=0.3\textwidth,angle=270]{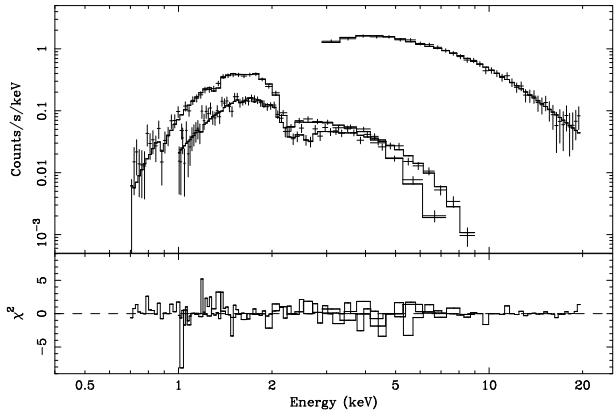}}
\vspace{0.2cm}
\caption{The observation 1 {\em Chandra} and PCA energy spectra 
fitted with an absorbed power-law model.  The {\em Chandra}
instruments used are the ACIS/HETG/MEG (0.7-7 keV) and the
ACIS/HETG/HEG (1-9 keV).  The measured power-law photon index 
is $\Gamma = 1.66\pm 0.05$.  The bottom panel shows the residuals.
\label{fig:spectrum1}}
\end{figure}

Using XSPEC 11.2, we performed simultaneous least-squares 
fits to the {\em Chandra} and PCA spectra for all three
observations.  For the PCA, we included 0.6\% and 0.3\% 
systematic errors for energy bins below and above 8~keV, 
respectively, as described in our previous work 
\citep{tomsick03}.  We began by fitting the observation 1 
spectrum with a power-law with interstellar absorption.  
We also included a constant factor to allow for the 
possibility that the overall normalization is different
for the PCA and {\em Chandra}, but we assumed that the 
overall MEG and HEG normalizations agree with each other. 
As shown in Table~\ref{tab:spec}, a power-law model with 
$\Gamma = 1.66\pm 0.05$ and $N_{\rm H} = (6.7\pm 0.5)\times 
10^{21}$ cm$^{-2}$ (errors are 90\% confidence, 
$\Delta\chi^{2} = 2.7$) provides a good fit to the spectrum
($\chi^{2}/\nu = 116/155$).  The other continuum models we 
tried do not provide acceptable fits.  Thermal models, such 
as a blackbody or a disk-blackbody \citep{makishima86}, 
result in reduced $\chi^{2}$ values of 8.4 and 2.3, 
respectively.  Figure~\ref{fig:spectrum1} shows the 
observation 1 spectrum and residuals with the power-law fit.
No clear features are seen in the residuals, and we note that 
the fit is not improved by adding a soft component, a high 
energy cutoff, or an iron emission line.  The upper limit 
on the equivalent width of a narrow (where ``narrow'' is set 
by the binning shown in Figure~\ref{fig:spectrum1}) Fe 
K$\alpha$ emission line at 6.4~keV is 161~eV.  

Table~\ref{tab:obs} and Figure~\ref{fig:lc} indicate that 
the count rates are very similar for observations 1 and 2.
We fitted the observation 2 spectrum, and Table~\ref{tab:spec} 
shows that the spectral parameters do not change significantly 
between the two observations.  Also, in both cases, the 
spectral fits indicate that the PCA/{\em Chandra} normalization 
factor is near 1.12.  As the PCA exposure only covers part of 
the {\em Chandra} observation, we checked to see if this 
difference could be due to source variability.  We extracted 
a grating spectrum for observation 1 that is strictly 
simultaneous with the PCA exposure.  During this time, the 
MEG count rate is $0.483\pm 0.016$ s$^{-1}$ (0.7-7 keV), which 
is consistent with the rate for the full observation ($0.479\pm 
0.007$ s$^{-1}$).  Thus, we conclude that the overall 
normalizations are different for the PCA and the {\em Chandra} 
gratings, and we note that this is not surprising since it 
has been previously determined that the PCA normalization 
is high relative to previous instruments \citep{tomsick99}.  

Due to the presence of photon pile-up in the PSF core, 
we extracted the observation 3 ACIS spectrum after 
removing the counts detected in the PSF core.  We
extracted the source counts from an elliptical annulus
centered on the XTE J1650--500 position.  We used
an inner ellipse with a semi-major axis of 
$1^{\prime\prime}.3$ and a semi-minor axis of 
$0^{\prime\prime}.9$, and an outer ellipse with a 
semi-major axis of $4^{\prime\prime}.9$ and a semi-minor 
axis of $3^{\prime\prime}.1$.  We rotated the ellipses 
to match the off-axis {\em Chandra} PSF.  This region 
contains 1573 counts compared to the 9000 counts 
obtained when the core is included (see Table~\ref{tab:obs}), 
and we are confident that the photons from the elliptical
annulus do not suffer from pile-up.  However, it is critical 
to realize that ACIS response matrices produced by the 
standard CIAO software do not account for the size or shape 
of the source extraction region or the fact that the 
{\em Chandra} PSF is energy dependent.  We corrected the 
ACIS response matrix for this using the {\em Chandra} Ray 
Tracing software (ChaRT) as described in Appendix B.  
When producing the response matrix we also applied the 
low energy ACIS response correction for detector evolution
as described above for the grating spectra.  

\begin{figure}
\centerline{\includegraphics[width=0.3\textwidth,angle=270]{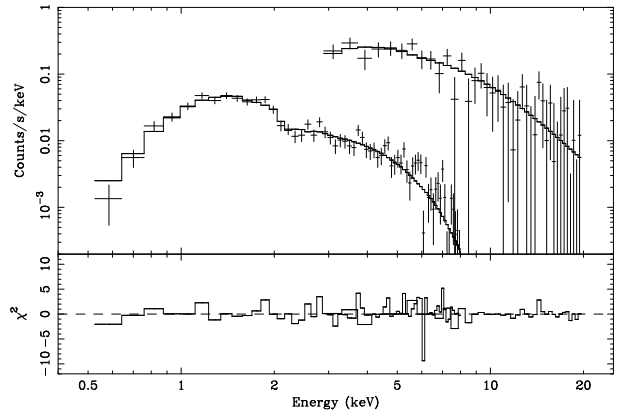}}
\vspace{0.2cm}
\caption{The observation 3 {\em Chandra} and PCA energy spectra 
fitted with an absorbed power-law model.  We obtained the 
{\em Chandra} spectrum (0.5-8 keV) with direct ACIS imaging.
The measured power-law photon index is $\Gamma = 1.93\pm 0.13$.
The bottom panel shows the residuals.
\label{fig:spectrum3}}
\end{figure}

We fitted the observation 3 ACIS+PCA spectrum with an 
absorbed power-law model as shown in Figure~\ref{fig:spectrum3}.  
This model provides a good fit to the data ($\chi^{2}/\nu = 
104/100$) with parameter values of $N_{\rm H} = (6.4\pm 
0.7)\times 10^{21}$ cm$^{-2}$ and $\Gamma = 1.93\pm 0.13$ 
(see Table~\ref{tab:spec}).  The measured column density is 
consistent with the value obtained from the grating observations.  
Thus, we re-fitted the spectrum after fixing the column density 
to the value found using the grating spectra ($N_{\rm H} = 
6.7\times 10^{21}$ cm$^{-2}$), and we obtain $\Gamma = 
1.96\pm 0.09$.  Also, for consistency with observations 
1 and 2, we used the PCA normalization for observation 3
divided by the PCA/grating normalization factor of 1.12
to determine the observation 3 flux reported in 
Table~\ref{tab:spec}.  The spectral results indicate that the 
XTE J1650--500 keeps its power-law shape but softens somewhat 
at low luminosity.

\subsection{Power Spectra at Low Flux Levels}

\begin{figure}
\centerline{\includegraphics[width=0.5\textwidth]{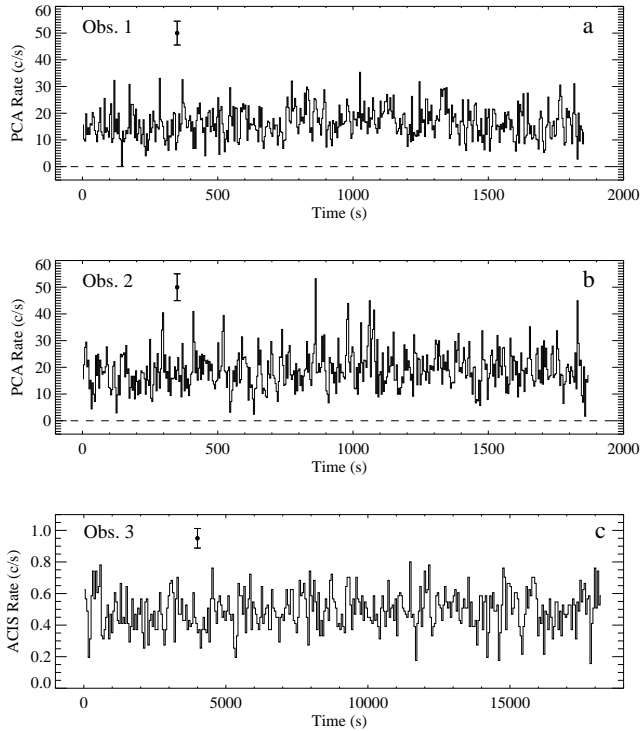}}
\vspace{0.2cm}
\caption{PCA (panels a and b) and ACIS (panel c) light curves
for the 3 observations.  The energy bands are 3-20 keV and
0.5-8 keV for the PCA and ACIS, respectively, and the light
curves have been background subtracted.  In each case, a 
representative error bar is shown.\label{fig:lc2}}
\end{figure}

We examined the PCA and {\em Chandra} light curves
for all three observations.  As shown in Figure~\ref{fig:lc2}, 
clear variability is present in the PCA light curves for 
observations 1 and 2 and in the ACIS light curve for 
observation 3.  The plotted 3-20 keV PCA light curves 
(panels a and b) have 4~s time bins, and we note that the 
count rate is higher than reported in Table~\ref{tab:obs} 
because we used all anode layers and PCUs 0, 2, and 3 
rather than just the top layer and PCUs 2 and 3.  
While the quality of the background estimate is slightly 
worse when PCU 0 and all layers are included, the higher 
count rate allows for tighter constraints on the timing
properties.  We also produced grating light curves (MEG+HEG) 
for observations 1 and 2.  Although the statistical quality 
of the PCA light curves is superior to that of the grating 
light curves, we use the grating light curves for the analysis 
below as they allow us to study the energy dependence of the 
timing properties.  Figure~\ref{fig:lc2}c shows the 0.5-8 keV 
ACIS light curve for observation 3 with $\sim$50~s time bins.  
We also examined the PCA light curve for observation 3, but it 
is background dominated and is not useful for timing analysis.  

\begin{figure}
\centerline{\includegraphics[width=0.5\textwidth]{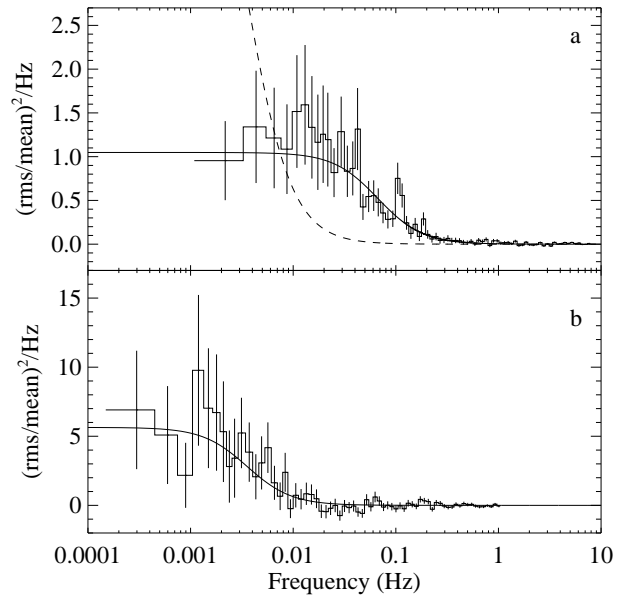}}
\vspace{0.2cm}
\caption{(a) PCA power spectrum for observations 1 and 2
combined.  The solid line shows the best fit zero-centered
Lorentzian, and the power spectrum shows marginal (between
2 and 3-$\sigma$) evidence for a QPO at 0.11~Hz.  (b) ACIS
power spectrum for observation 3 with the best fit 
zero-centered Lorentzian.  In panel a, the dashed line 
shows the model used for observation 3 to illustrate that
the power shifts to lower frequency for observation 3.
\label{fig:power}}
\end{figure}

For observations 1 and 2, we produced power spectra 
using PCA data taken in a high time resolution ``event''
mode (labeled E\_125us\_64M\_0\_1s).  These data consist
of an event list with photons time tagged to within
125 $\mu$s and 64 energy bins.  First, we used the
data in the 3-20 keV band to produce Leahy normalized 
power spectra \citep{leahy83} for observations 1 and 2 
separately.  In each case, we averaged power spectra 
from four 460~s light curve segments and used a 
Nyquist frequency of 64~Hz.  No noise in excess of 
the Poisson level is present above $\sim$1~Hz, and
there were no clearly significant differences 
between the power spectra for the two observations.
Thus, we produced another Leahy normalized power 
spectrum (again with a 3-20 keV energy band) consisting 
of the eight 460~s light curve segments from 
observations 1 and 2 with a Nyquist frequency of 8~Hz.  
We converted to the rms normalization \citep{miyamoto94}, 
and this power spectrum is shown in Figure~\ref{fig:power}a.  
The power spectrum is dominated by a band-limited noise 
component, and such a component has typically been fitted 
with a broken power-law or a zero-centered Lorentzian in 
previous work \citep{belloni02}.  A zero-centered Lorentzian 
with a half-width (FWHM/2) of $0.067\pm 0.007$~Hz, and a 
fractional rms amplitude of 33.1\%$\pm 1.1$\% (see 
Table~\ref{tab:power}) provides a reasonably good fit to 
the power spectrum ($\chi^{2}/\nu = 93/72$).  We note 
that the half-width of a zero-centered Lorentzian is
equivalent to the ``peak'' frequency that is sometimes 
quoted when fitting Lorentzians to power spectra 
\citep{belloni02,kalemci03b}.  Positive residuals 
are present near 0.1~Hz, possibly indicating the presence 
of a quasi-periodic oscillation (QPO).  The addition of a 
second Lorentzian at 0.11~Hz improves the quality of the fit, 
resulting in $\chi^{2}/\nu = 83/69$; however, this improvement 
corresponds to a detection at only the 96\% confidence level.  
Although we repeated this analysis several times with different 
frequency binnings, we consistently found that the QPO is only 
detected at a level of between 2 and 3-$\sigma$.  Thus, we 
take the measured rms amplitude of 12\% as an upper limit on 
a possible QPO at 0.11~Hz.

Using the grating light curves, we produced a 0.7-9 keV 
power spectrum for observations 1 and 2 by averaging 
eight 2414~s light curve segments.  The time resolution 
is set by the interval of time between chip read-outs, and 
this interval is 2.54104~s for these observations, giving
a Nyquist frequency of 0.2~Hz.  Although the time resolution 
and the statistical quality are lower for the grating light 
curves than for the PCA data, using the grating data allows 
us to extend the timing study to lower energies and frequencies.
After converting to rms normalization, we fitted the grating
power spectrum with a zero-centered Lorentzian, and the fit
is acceptable ($\chi^{2}/\nu = 25/23$).  As shown in 
Table~\ref{tab:power}, the half-width is $0.079\pm 0.020$~Hz, 
and the fractional rms is 30.1\%$\pm 2.5$\%.  Thus, the 
parameters for the 0.7-9 keV power spectrum are consistent 
with those for the 3-20 keV PCA power spectrum.

We also produced a power spectrum using the observation 
3 ACIS data.  Using a 0.5-8 keV light curve with a time 
resolution of 0.44104~s, we produced a Leahy normalized 
power spectrum with a Nyquist frequency of 1.13~Hz, 
consisting of the average of six 3350~s light curve 
intervals.  In the Leahy normalization, Poisson noise 
leads to a power level of 2.0; however, at high frequencies 
(0.1-1~Hz) the ACIS power spectrum is consistent with a 
constant level of 1.77.  We performed simulations described 
in Appendix A showing that pile-up at the level present for 
observation 3 leads to a drop in the Poisson noise level 
from 2.0 to 1.78.  This, along with a small drop in the 
Poisson noise level caused by deadtime during chip read-outs 
(see Appendix A) explains our measurement of 1.77.  Thus, 
when converting the ACIS power spectrum from the Leahy 
normalization to the rms normalization, we subtracted 
a value of 1.77 rather than the usual value of 2.0.  The 
rms normalized power spectrum fitted with a zero-centered 
Lorentzian is shown in Figure~\ref{fig:power}b, and this 
model provides a good fit to the data ($\chi^{2}/\nu = 56/65$).
The Lorentzian half-width is $0.0035\pm 0.0010$~Hz (see 
Table~\ref{tab:power}), which is significantly lower than 
the value obtained for observations 1 and 2.  This indicates 
that, as the flux drops, the band-limited noise moves to lower 
frequencies.  This fact is illustrated in Figure~\ref{fig:power}a, 
where the model used to fit the observation 3 power spectrum is 
plotted as a dashed line on the power spectrum for observations 
1 and 2.  The fit to the observation 3 power spectrum indicates 
a fractional rms amplitude of 17.7\%$\pm 1.7$\%, but this (16\%)
should be taken as a lower limit on the actual value since the 
noise level is reduced by pile-up.  From our simulations, 
we estimate that the actual rms amplitude is near 19\%.  We 
also produced a power spectrum using only the 1573 counts from 
the region of the CCD that is free of pile-up.  Fitting the 
power spectrum with a zero-centered Lorentzian, the half-width 
is not well-constrained, but it is consistent with the previous
fit, and the rms amplitude is 19\%$\pm 7$\%, giving an upper 
limit of 26\%.  Thus, considering the fits to the power 
spectra with and without pile-up, we conclude that the rms 
is 19\%$^{+7\%}_{-3\%}$.

\subsection{Summary of XTE J1650--500 Hard State X-Ray Observations}

For observations 1-3, two of our main results are that the source 
softens ($\Gamma$ changes from $1.66\pm 0.05$ to $1.93\pm 0.13$) 
and the band-limited noise component moves to lower frequencies 
(the Lorentzian half-width, FWHM/2, changes from $0.067\pm 
0.007$~Hz to $0.0035\pm 0.0010$~Hz) as the source flux decays.  
Thus, where possible, we determined the values of these parameters 
($\Gamma$ and FWHM/2) for other {\em RXTE} observations of
XTE J1650--500 in the hard state during outburst decay.  For the
7 observations made immediately after the soft-to-hard state
transition on MJD 52231.5 (see Figure~\ref{fig:lc}), we use the
spectral parameters from \cite{kalemci_thesis}.  These energy
spectra are dominated by a power-law component, but it should
be noted that a soft component is also present.  For these 7
observations, the timing parameters come from \cite{kalemci03b}.
In contrast to our {\em Chandra}/{\em RXTE} observations at low 
flux levels, several Lorentzians are required to describe the 
power spectra for most of the observations immediately after the 
soft-to-hard state transition.  The lowest frequency component
in those power spectra are nearly zero-centered Lorentzians, 
and we assume that they correspond to the band-limited noise
component that we see in the low flux power spectra.

\begin{figure}
\centerline{\includegraphics[width=0.5\textwidth]{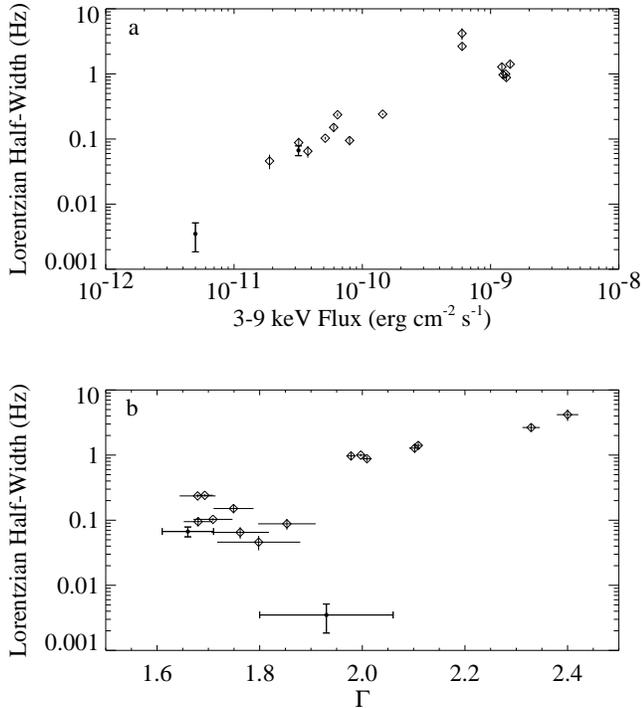}}
\vspace{-0.1cm}
\caption{Measured parameters from {\em Chandra} and {\em RXTE}
observations of XTE J1650--500 in the hard state.  (a) The 
values of the Lorentzian half-width vs. the flux in the 3-9 keV 
energy band. (b) The half-width vs. the power-law photon index 
($\Gamma$).  In both panels, the diamonds mark the measurements 
from observations with only {\em RXTE} data and the filled
circles mark observations where we obtained both {\em Chandra}
and {\em RXTE} data (one point for observations 1+2 and one
point for observation 3).  The error bars on frequency and
on $\Gamma$ are 90\% confidence.\label{fig:summary}}
\end{figure}

For the 15 {\em RXTE} observations after the gap in {\em RXTE} 
coverage and before the {\em Chandra} observations began, 
the 3-20 keV energy spectra are all well-described by a single 
power-law component with the column density fixed to the value 
of $N_{\rm H} = 6.7\times 10^{21}$ cm$^{-2}$ found for the 
grating observations.  For these observations, the 3-20 keV 
flux varies from $1.2\times 10^{-11}$ to $2.9\times 10^{-10}$ 
erg~cm$^{-2}$~s$^{-1}$, and the values of $\Gamma$ range 
from $1.69\pm 0.02$ to $\sim$2.0.  The data suggest a trend 
of softening as the flux drops consistent with that observed 
with {\em Chandra}, but the errors on $\Gamma$ are large 
for the low flux observations.  We produced power spectra 
for all 15 observations; however, we could not obtain reliable 
fit parameters for about half of the observations due to poor 
statistics, short observations, or flaring \citep{tomsick03}.  
In 8 cases, we obtained useful power spectra, and these are 
dominated by a band-limited component with a fractional rms 
amplitude near 30\%.  We fitted these power spectra with a 
zero-centered Lorentzian to determine the half-width for 
comparison to the other observations.  Figure~\ref{fig:summary}a 
shows the measured values of the Lorentzian half-width vs. 
3-9 keV flux, and Figure~\ref{fig:summary}b shows the 
half-width vs. the measured values of $\Gamma$.  

For the data points above a half-width of 0.5~Hz, there
is a tight correlation between $\Gamma$ and the half-width 
with a linear correlation coefficient of 0.96.  The two
points near 0.25~Hz also appear to be consistent with 
this correlation, but the correlation coefficient drops 
slightly to 0.91 when these points are included.  It
is clear that the correlation does not extend to lower
frequencies as $\Gamma$ reaches a minimum near 1.65-1.70.  
In fact, the data at frequencies between 0.03~Hz and 0.3~Hz 
suggest the possibility of an anti-correlation between 
$\Gamma$ and the half-width, but the linear correlation 
coefficient is only --0.41.  We checked to see if it is
valid to compare the {\em Chandra}/{\em RXTE} point in 
this frequency range to the {\em RXTE} only points by 
fitting the {\em RXTE} spectra alone for observations 
1 and 2.  Fixing $N_{\rm H}$ to $6.7\times 10^{21}$ cm$^{-2}$, 
we obtained values of $\Gamma$ consistent with those found 
using {\em Chandra} and {\em RXTE} together.  Although the 
anti-correlation is not statistically significant, the 
distribution of the values of $\Gamma$ for the 0.03-0.3 Hz 
points is also not well-described by a constant at their 
weighted average of $\Gamma = 1.70\pm 0.01$.  Thus, we 
conclude that there is some scatter in $\Gamma$ for the 
0.03-0.3 Hz points, but we cannot establish a definite 
relationship between $\Gamma$ and the Lorentzian half-width 
in this frequency regime.

\section{Discussion}

One of the main motivations for this work is to measure 
the X-ray properties of an accreting black hole in the 
relatively unexplored luminosity range between the 
transition to the low-hard (or hard) state and the
quiescent state.  Such measurements allow us to address
the question of whether these should be considered as
two separate states or if the quiescent state is, in 
fact, a low luminosity version of the hard state.  This
question has important implications for whether there
are significant physical changes in accreting black 
hole systems that occur at low luminosities such as 
changes in the geometry of the accretion flow and/or 
emission mechanisms.

For XTE J1650--500, we have measured the spectral and 
timing properties of the source at 1-9 keV (absorbed) 
luminosities of $9\times 10^{34}$ erg~s$^{-1}$ 
(observations 1 and 2) and $1.5\times 10^{34}$ erg~s$^{-1}$ 
(observation 3).  The quoted luminosities assume a fiducial 
source distance of 4~kpc, and we note that the unabsorbed 
luminosities are about 25\% higher.  As discussed in 
\cite{tomsick03}, the likely distance for XTE J1650--500 
is between 2~kpc and 6~kpc.  For the {\em RXTE} observation 
immediately after the soft-to-hard state transition, we 
used the spectral model described in \cite{kalemci_thesis} 
to extrapolate the PCA-measured flux into the 1-9 keV 
energy band, and we infer a transition luminosity of 
$7\times 10^{36}$ erg~s$^{-1}$, which is in-line with 
typical transition luminosities \citep{ts96}.  The 
observation 3 luminosity, which is a factor of 470 lower 
than the transition flux, is only about an order of 
magnitude above the quiescent luminosity of the black 
hole system that is brightest in quiescence (V404~Cyg, 
Garcia et al. 2001\nocite{garcia01}).  Thus, our 
XTE J1650--500 observations cover a significant portion 
of the luminosity range between the soft-to-hard state 
transition and quiescence, but we note that the quiescent 
luminosity for XTE J1650--500 could be considerably lower 
than for V404~Cyg.

The trend observed for XTE J1650--500 that the band-limited 
noise gradually shifts to lower frequency immediately after 
the transition to the hard state \citep{kalemci03b} has 
also been observed for several other black hole systems
\citep{kalemci03a,nwd02}.  The drop in frequency can be 
explained by accretion models consisting of an inner 
optically thin corona and an outer optically thick disk.  
As the mass accretion rate drops, the inner radius of the 
optically thick disk ($R_{in}$) moves away from the black 
hole, causing a drop in the dynamical time scales and, thus, 
the characteristic frequencies for the system \citep{dp99}.
After the XTE J1650--500 transition to the hard state, 
the Lorentzian half-width is correlated with the value of 
the power-law index down to at least a half-width of 1~Hz 
and possibly down to 0.25~Hz (see \S 3.3).  Correlations 
between frequency (either half-width or QPO frequency) and 
$\Gamma$ have been seen before during {\em RXTE} observations 
of Cyg X-1, GX 339--4, and other black hole systems in their 
hard states \citep{gcr99,rgc01,kalemci_thesis}.  As these 
authors argue, this correlation is expected for the disk 
and inner corona picture because the soft photons from the 
optically thick disk cool the corona, leading to a softer 
power-law index.  Thus, as $R_{in}$ increases, and the 
characteristic frequencies drop, the corona is subject to 
a lower flux of soft photons from the disk, and the coronal 
spectrum hardens.  This effect is predicted in both ADAF 
\citep{emn97} models and in general disk and inner corona 
models such as the sphere+disk model \citep{nwd02} where the 
X-ray emission from the corona is predominantly due to thermal 
Comptonization.

While the half-width and $\Gamma$ are correlated above
0.25-1~Hz, $\Gamma$ does not get any harder than 1.65-1.70
for XTE J1650--500, and the spectrum softens to $1.93\pm 0.13$ 
as the half-width and flux continue to drop.  This behavior can 
be interpreted within the ADAF or sphere+disk (henceforth ADAF/SD) 
models.  The decrease in the half-width frequency indicates 
that the disk inner radius continues to recede; however, the 
disk flux is at such a low level that it no longer significantly 
cools the corona, and the spectrum does not continue to get harder.  
Instead, the hardness of the spectrum depends on the 
``Compton y-parameter,'' which is a function of the optical depth 
and the temperature of the corona \citep{rl79}.  As the mass 
accretion rate drops, the coronal optical depth decreases, and 
a softer spectrum is predicted.  Softer spectra have been observed 
at low luminosities for several black hole sources besides 
XTE J1650--500.  XTE J1118+480 shows significant softening that 
starts at an X-ray luminosity of $2\times 10^{35}$ erg~s$^{-1}$ 
($d$/1.8 kpc)$^{2}$, and, at the end of its 2001 outburst, 
XTE J1550--564 begins gradual softening at a luminosity 
of $2\times 10^{36}$ erg~s$^{-1}$ ($d$/5.3 kpc)$^{2}$
\citep{kalemci_thesis}.  Other examples of spectral softening
at low luminosities include GS 1124--68 \citep{ebisawa94}, 
as well as some black hole sources in or close to quiescence 
\citep{mcclintock03,corbel03,kong02a,tck01}.

Although spectral fits with the ADAF model imply relatively 
large changes in $R_{in}$ between the soft and hard states, 
some previous black hole timing studies conclude that such
changes, which can be as large as factors of 30-3000 
\citep{emn97}, do not occur.  For several black hole systems, 
\cite{dp99} find that the characteristic frequencies do not 
vary by more than a factor of 50 between states.  For 
Keplerian orbits, $R_{in}\propto \nu^{-2/3}$, where $\nu$ 
is the characteristic frequency, so that a factor of 50 in 
frequency implies a change in $R_{in}$ by a factor of 14, 
which is significantly less than indicated by ADAF fits.  
However, for XTE J1650--500, we find that the Lorentzian 
half-width changes by a factor of nearly 1200 while the 
source remains in the hard state, implying that $R_{in}$ can 
change by as much as a factor of 113.  Results of fast optical 
photometry for quiescent black hole systems suggest that 
$R_{in}$ may change by an even larger factor.  For A~0620--00, 
\cite{hynes03} find that the optical power spectrum is 
dominated by band limited noise with a break frequency of 
$9.5\times 10^{-4}$ Hz, which is a factor of 3.7 lower than 
the lowest frequency we find for XTE J1650--500.  If the 
frequency is, in fact, related to $R_{in}$, then our result 
for XTE J1650--500 and the results of \cite{hynes03} imply 
that very large values of $R_{in}$ occur at low luminosities, 
and that the $R_{in}$ evolution occurs gradually and 
continues to change all the way down to quiescent luminosities.
These results also imply that an extremely large value of $R_{in}$ 
is not required for a source to show hard state properties.

While the ADAF/SD models described above can explain many 
of the X-ray properties we see for XTE J1650--500, radio 
observations of XTE J1650--500 made after the transition to 
the hard state indicate the likely presence of a compact jet 
(Corbel et al., in prep.), which is not accounted for in 
these models.  However, the ``magnetic corona'' model of 
\cite{mf02} also explains several of the XTE J1650--500 
X-ray properties while predicting that energy stored in the 
corona should be available to drive strong outflows.  In 
this model, the X-ray emission mechanism is also 
Comptonization, so that a high flux of disk photons leads 
to a softer energy spectrum.  Thus, \cite{mf02} predict that 
the spectrum will harden as the mass accretion rate drops, 
and they also predict that the spectrum will soften at very 
low mass accretion rates as more of the coronal energy is 
carried away by the outflow.  Recent work by \cite{fgj03} 
suggests that we are observing XTE J1650--500 in a 
luminosity regime where this could occur.  While the 
XTE J1650--500 spectral evolution is consistent with the 
\cite{mf02} model, and we also previously argued that this 
model may be able to account for the X-ray flares reported 
in \cite{tomsick03}, it is unclear if the magnetic corona 
model can account for the fact that the observed characteristic 
frequency for XTE J1650--500 changes by a factor of 1200 in 
the hard state.  Perhaps this could indicate large variations 
in the size of the magnetic corona or in the sizes of the 
active coronal regions, but since $R_{in}$ is not expected 
to change by a large factor in this model, we do not see an 
obvious explanation for the large frequency range.

A jet-dominated model has also been suggested for
black holes in their hard state.  In this model, the
compact jet contributes a large fraction of the emission
from the radio band up to X-ray energies, and a
synchrotron emission mechanism is invoked \citep{mff01}.
Evidence in favor of this model includes observations
of correlated radio and X-ray flux for several black 
hole systems \citep{corbel03,gfp03}, and good agreement
between the predictions of the model and the 
broadband GX 339--4 energy spectra over a large range
of mass accretion rates \citep{markoff03}.  In the
model, X-rays are produced when electrons are 
accelerated at a shock that is $z_{sh}\sim 10^{3} R_{g}$
from the black hole, where $R_{g} = GM/c^{2}$ and 
$M$ is the black hole mass.  The X-ray spectrum is
predicted to have a power-law shape out to a cutoff
energy, and the power-law index is set by the index
of the electron energy distribution, $p$, where 
$dN/dE \propto E^{-p}$.  \cite{markoff03} find that
they can explain the radio/X-ray flux correlation
for GX 339--4 by varying only the jet power, 
keeping $p$ and $z_{sh}$ constant.  The fact that
$\Gamma$ changes for XTE J1650--500 and also for
GX 339--4 \citep{corbel03}, indicates that if this
model is correct, $p$ likely does change with source
luminosity, but we suspect that changes in $\Gamma$
are probably not a major problem for this model
as one might expect $p$ to vary somewhat with jet
power.  On the other hand, one would expect that 
the characteristic frequencies of the system would 
be related to the size of the source.  Assuming
that the size of the X-ray emitting region is 
related to the location of the shock, $z_{sh}$, 
it may be difficult for the jet-dominated model
to explain the large range of characteristic
frequencies we observe for XTE J1650--500 without
varying $z_{sh}$.

\section{Conclusions}

We have compared the emission properties of XTE J1650--500 
in the hard state over a large range of luminosities to 
ADAF/SD, magnetic corona, and jet-dominated models.  
Although the ADAF/SD model provides natural explanations 
for the X-ray spectral and timing properties, these models 
do not include outflows and thus cannot provide a complete 
description of the system.  In addition, ADAFs may be 
convectively unstable as mentioned above.  The magnetic 
corona model, which does incorporate a jet, predicts a 
similar spectral evolution to the ADAF/SD model, but it it 
unclear if it can explain the large change in the Lorentzian 
half-width that we see for XTE J1650--500.  The jet-dominated 
model can explain X-ray/radio correlations that are seen for 
several black hole sources without varying $z_{sh}$; however, 
it is doubtful that the model can reproduce the observed X-ray 
timing properties with a constant value of $z_{sh}$.  

Regardless of which model is correct, the fact that 
the general shape of the energy and power spectra remain 
the same while the parameters evolve indicates that the 
X-ray emitting region of the system likely keeps the same 
overall structure even though the parameter changes 
indicate that some physical properties change, such as 
the inner radius of the accretion disk, the size of the 
corona, or the shock location, considering the models 
described above.  It is clear that the source retains 
its hard state X-ray properties down to a luminosity of 
$1.5\times 10^{34}$ erg~s$^{-1}$, and we do not see any 
evidence for a sharp transition to the quiescent state.  
Furthermore, our results for XTE J1650--500 combined
with the determination that the optical power spectrum
for A~0620--00 in quiescence exhibits band limited
noise with an even lower characteristic frequency 
\citep{hynes03} suggests that black hole transients 
keep the same overall structure into quiescence.
Thus, these results favor the hypothesis that the
quiescent state is simply a low luminosity version 
of the hard state.

\acknowledgements

JAT acknowledges useful discussions with M. A. Nowak, J. Wilms, 
and W. A. Heindl.  We thank M. Garcia and H. Tananbaum for 
assistance in scheduling the Target of Opportunity observations 
with {\em Chandra}.  We would like to thank all scientists who 
contributed to the T\"{u}bingen Timing Tools.  JAT acknowledges 
partial support from NASA grant NAG5-13055 and {\em Chandra} 
award number GO2-3056X issued by the {\em Chandra} X-ray 
Observatory Center, which is operated by the Smithsonian 
Astrophysical Observatory for and on behalf of NASA under 
contract NAS8-39073.  EK acknowledges partial support of 
T\"UB\.TAK.  PK acknowledges partial support from NASA grant 
NAG5-7405.


\clearpage

\begin{thebibliography}{}

\bibitem[\protect\astroncite{{Angelini} \& {White}}{2003}]{aw03}
{Angelini}, L., \& {White}, N.~E.,  2003, ApJ, 586, L71

\bibitem[\protect\astroncite{{Belloni}, {Psaltis} \& {van der
  Klis}}{2002}]{belloni02}
{Belloni}, T., {Psaltis}, D., \& {van der Klis}, M.,  2002, ApJ, 572, 392

\bibitem[\protect\astroncite{{Bradt}, {Rothschild} \& {Swank}}{1993}]{brs93}
{Bradt}, H.~V., {Rothschild}, R.~E., \& {Swank}, J.~H.,  1993, A\&AS, 97, 355

\bibitem[\protect\astroncite{{Castro-Tirado} et~al.}{2001}]{ct01}
{Castro-Tirado}, A.~J., {Kilmartin}, P., {Gilmore}, A., {Petterson}, O.,
  {Bond}, I., {Yock}, P., \& {Sanchez-Fernandez}, C.,  2001, IAU~Circular, 7707

\bibitem[\protect\astroncite{{Corbel} et~al.}{2000}]{corbel00}
{Corbel}, S., {Fender}, R.~P., {Tzioumis}, A.~K., {Nowak}, M., {McIntyre}, V.,
  {Durouchoux}, P., \& {Sood}, R.,  2000, A\&A, 359, 251

\bibitem[\protect\astroncite{{Corbel} et~al.}{2002}]{corbel02}
{Corbel}, S., {Fender}, R.~P., {Tzioumis}, A.~K., {Tomsick}, J.~A., {Orosz},
  J.~A., {Miller}, J.~M., {Wijnands}, R., \& {Kaaret}, P.,  2002, Science, 298,
  196

\bibitem[\protect\astroncite{{Corbel} et~al.}{2003}]{corbel03}
{Corbel}, S., {Nowak}, M.~A., {Fender}, R.~P., {Tzioumis}, A.~K., \& {Markoff},
  S.,  2003, A\&A, 400, 1007

\bibitem[\protect\astroncite{{Davis}}{2001}]{davis01}
{Davis}, J.~E.,  2001, ApJ, 562, 575

\bibitem[\protect\astroncite{{di Matteo}, {Celotti} \& {Fabian}}{1999}]{dcf99}
{di Matteo}, T., {Celotti}, A., \& {Fabian}, A.~C.,  1999, MNRAS, 304, 809

\bibitem[\protect\astroncite{{di Matteo} \& {Psaltis}}{1999}]{dp99}
{di Matteo}, T., \& {Psaltis}, D.,  1999, ApJ, 526, L101

\bibitem[\protect\astroncite{{Ebisawa} et~al.}{1994}]{ebisawa94}
{Ebisawa}, K., et~al., 1994, PASJ, 46, 375

\bibitem[\protect\astroncite{{Esin}, {McClintock} \& {Narayan}}{1997}]{emn97}
{Esin}, A.~A., {McClintock}, J.~E., \& {Narayan}, R.,  1997, ApJ, 489, 865

\bibitem[\protect\astroncite{Fender, Gallo \& Jonker}{2003}]{fgj03}
Fender, R., Gallo, E., \& Jonker, P.,  2003, Accepted by MNRAS,
  astro-ph/0306614

\bibitem[\protect\astroncite{{Fender}}{2001}]{fender01}
{Fender}, R.~P.,  2001, MNRAS, 322, 31

\bibitem[\protect\astroncite{{Galeev}, {Rosner} \& {Vaiana}}{1979}]{grv79}
{Galeev}, A.~A., {Rosner}, R., \& {Vaiana}, G.~S.,  1979, ApJ, 229, 318

\bibitem[\protect\astroncite{{Gallo}, {Fender} \& {Pooley}}{2003}]{gfp03}
{Gallo}, E., {Fender}, R.~P., \& {Pooley}, G.~G.,  2003, MNRAS, 344, 60

\bibitem[\protect\astroncite{{Garcia} et~al.}{2001}]{garcia01}
{Garcia}, M.~R., {McClintock}, J.~E., {Narayan}, R., {Callanan}, P., {Barret},
  D., \& {Murray}, S.~S.,  2001, ApJ, 553, L47

\bibitem[\protect\astroncite{{Gilfanov}, {Churazov} \&
  {Revnivtsev}}{1999}]{gcr99}
{Gilfanov}, M., {Churazov}, E., \& {Revnivtsev}, M.,  1999, A\&A, 352, 182

\bibitem[\protect\astroncite{{Groot} et~al.}{2001}]{groot01}
{Groot}, P., {Tingay}, S., {Udalski}, A., \& {Miller}, J.,  2001, IAU~Circular,
  7708

\bibitem[\protect\astroncite{{Homan} et~al.}{2003}]{homan03}
{Homan}, J., {Klein-Wolt}, M., {Rossi}, S., {Miller}, J.~M., {Wijnands}, R.,
  {Belloni}, T., {van der Klis}, M., \& {Lewin}, W.~H.~G.,  2003, ApJ, 586,
  1262

\bibitem[\protect\astroncite{{Hynes} et~al.}{2003}]{hynes03}
{Hynes}, R.~I., {Charles}, P.~A., {Casares}, J., {Haswell}, C.~A., {Zurita},
  C., \& {Shahbaz}, T.,  2003, MNRAS, 340, 447

\bibitem[\protect\astroncite{Kalemci}{2002}]{kalemci_thesis}
Kalemci, E.,  2002, Ph.D. Thesis, Temporal Studies of Black Hole X-Ray
  Transients During Outburst Decay, UC San Diego

\bibitem[\protect\astroncite{Kalemci et~al.}{2003a}]{kalemci03a}
Kalemci, E., et~al., 2003a, Submitted to ApJ

\bibitem[\protect\astroncite{{Kalemci} et~al.}{2003b}]{kalemci03b}
{Kalemci}, E., {Tomsick}, J.~A., {Rothschild}, R.~E., {Pottschmidt}, K.,
  {Corbel}, S., {Wijnands}, R., {Miller}, J.~M., \& {Kaaret}, P.,  2003b, ApJ,
  586, 419

\bibitem[\protect\astroncite{{Kong} et~al.}{2000}]{kong00}
{Kong}, A. K.~H., {Kuulkers}, E., {Charles}, P.~A., \& {Homer}, L.,  2000,
  MNRAS, 312, L49

\bibitem[\protect\astroncite{{Kong} et~al.}{2002}]{kong02a}
{Kong}, A.~K.~H., {McClintock}, J.~E., {Garcia}, M.~R., {Murray}, S.~S., \&
  {Barret}, D.,  2002, ApJ, 570, 277

\bibitem[\protect\astroncite{{Leahy} et~al.}{1983}]{leahy83}
{Leahy}, D.~A., {Darbro}, W., {Elsner}, R.~F., {Weisskopf}, M.~C., {Kahn}, S.,
  {Sutherland}, P.~G., \& {Grindlay}, J.~E.,  1983, ApJ, 266, 160

\bibitem[\protect\astroncite{{Makishima} et~al.}{1986}]{makishima86}
{Makishima}, K., {Maejima}, Y., {Mitsuda}, K., {Bradt}, H.~V., {Remillard},
  R.~A., {Tuohy}, I.~R., {Hoshi}, R., \& {Nakagawa}, M.,  1986, ApJ, 308, 635

\bibitem[\protect\astroncite{{Markoff}, {Falcke} \& {Fender}}{2001}]{mff01}
{Markoff}, S., {Falcke}, H., \& {Fender}, R.,  2001, A\&A, 372, L25

\bibitem[\protect\astroncite{{Markoff} et~al.}{2003}]{markoff03}
{Markoff}, S., {Nowak}, M., {Corbel}, S., {Fender}, R., \& {Falcke}, H.,  2003,
  A\&A, 397, 645

\bibitem[\protect\astroncite{McClintock \& Remillard}{2003}]{mr03}
McClintock, J., \& Remillard, R.,  2003, Review Article, astro-ph/0306213

\bibitem[\protect\astroncite{{McClintock} et~al.}{2003}]{mcclintock03}
{McClintock}, J.~E., {Narayan}, R., {Garcia}, M.~R., {Orosz}, J.~A.,
  {Remillard}, R.~A., \& {Murray}, S.~S.,  2003, ApJ, 593, 435

\bibitem[\protect\astroncite{{Merloni} \& {Fabian}}{2002}]{mf02}
{Merloni}, A., \& {Fabian}, A.~C.,  2002, MNRAS, 332, 165

\bibitem[\protect\astroncite{{Miyamoto} et~al.}{1994}]{miyamoto94}
{Miyamoto}, S., {Kitamoto}, S., {Iga}, S., {Hayashida}, K., \& {Terada}, K.,
  1994, ApJ, 435, 398

\bibitem[\protect\astroncite{{Narayan}, {McClintock} \& {Yi}}{1996}]{nmy96}
{Narayan}, R., {McClintock}, J.~E., \& {Yi}, I.,  1996, ApJ, 457, 821

\bibitem[\protect\astroncite{{Nowak} et~al.}{1999}]{nowak99}
{Nowak}, M.~A., {Vaughan}, B.~A., {Wilms}, J., {Dove}, J.~B., \& {Begelman},
  M.~C.,  1999, ApJ, 510, 874

\bibitem[\protect\astroncite{{Nowak}, {Wilms} \& {Dove}}{2002}]{nwd02}
{Nowak}, M.~A., {Wilms}, J., \& {Dove}, J.~B.,  2002, MNRAS, 332, 856

\bibitem[\protect\astroncite{{Quataert} \& {Gruzinov}}{2000}]{qg00}
{Quataert}, E., \& {Gruzinov}, A.,  2000, ApJ, 539, 809

\bibitem[\protect\astroncite{{Remillard}}{2001}]{remillard01}
{Remillard}, R.,  2001, IAU~Circular, 7707

\bibitem[\protect\astroncite{{Revnivtsev}, {Gilfanov} \&
  {Churazov}}{2001}]{rgc01}
{Revnivtsev}, M., {Gilfanov}, M., \& {Churazov}, E.,  2001, A\&A, 380, 520

\bibitem[\protect\astroncite{{Rybicki} \& {Lightman}}{1979}]{rl79}
{Rybicki}, G.~B., \& {Lightman}, A.~P.,  1979,
\newblock {Radiative Processes in Astrophysics},
\newblock  New York, Wiley-Interscience, 1979.~393 p.)

\bibitem[\protect\astroncite{{Sanchez-Fernandez} et~al.}{2002}]{sf02}
{Sanchez-Fernandez}, C., {Zurita}, C., {Casares}, J., {Castro-Tirado}, A.~J.,
  {Bond}, I., {Brandt}, S., \& {Lund}, N.,  2002, IAU~Circular, 7989

\bibitem[\protect\astroncite{{Tanaka} \& {Shibazaki}}{1996}]{ts96}
{Tanaka}, Y., \& {Shibazaki}, N.,  1996, ARA\&A, 34, 607

\bibitem[\protect\astroncite{Tomsick et~al.}{2003}]{tomsick03}
Tomsick, J., Corbel, S., Kalemci, E., \& Kaaret, P.,  2003, ApJ, 592, 1100

\bibitem[\protect\astroncite{{Tomsick}, {Corbel} \& {Kaaret}}{2001}]{tck01}
{Tomsick}, J.~A., {Corbel}, S., \& {Kaaret}, P.,  2001, ApJ, 563, 229

\bibitem[\protect\astroncite{{Tomsick} \& {Kaaret}}{2000}]{tk00}
{Tomsick}, J.~A., \& {Kaaret}, P.,  2000, ApJ, 537, 448

\bibitem[\protect\astroncite{{Tomsick} et~al.}{1999}]{tomsick99}
{Tomsick}, J.~A., {Kaaret}, P., {Kroeger}, R.~A., \& {Remillard}, R.~A.,  1999,
  ApJ, 512, 892

\bibitem[\protect\astroncite{{Valinia} \& {Marshall}}{1998}]{vm98}
{Valinia}, A., \& {Marshall}, F.~E.,  1998, ApJ, 505, 134

\bibitem[\protect\astroncite{{Weisskopf} et~al.}{2002}]{weisskopf02}
{Weisskopf}, M.~C., {Brinkman}, B., {Canizares}, C., {Garmire}, G., {Murray},
  S., \& {Van Speybroeck}, L.~P.,  2002, PASP, 114, 1

\end{thebibliography}

\begin{center}
{\small APPENDIX A}
\end{center}

We performed simulations to study the effects of pile-up on 
the observation 3 ACIS power spectrum.  The simulations have 
3 main goals:  1.  To understand why the Poisson noise level 
for the actual Leahy normalized ACIS power spectrum is 1.77; 
2. To determine how much the fractional rms amplitude is 
reduced by pile-up; and 3. To check that pile-up does not 
alter the overall shape of the power spectrum.  Our goals 
were accomplished via 3 simulations.

\begin{figure}
\centerline{\includegraphics[width=0.5\textwidth]{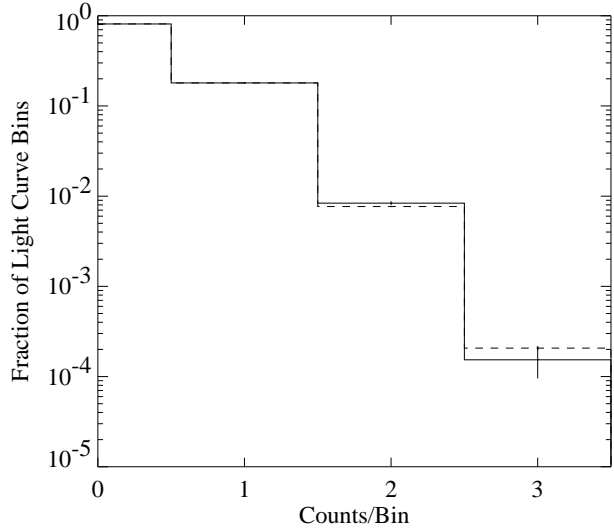}}
\vspace{0.2cm}
\caption{The solid histogram shows the fraction of 
0.4~s observation 3 light curve bins with with 0, 1, 
2, or 3 counts/bin.  The dashed histogram shows the 
same for a simulated light curve that includes the 
effects of photon pile-up.  This illustrates
that our simulated light curves are a good match
to the actual data.\label{fig:pileup}}
\end{figure}

Simulation \#1:  We produced a simulated Poisson noise light 
curve with the same time resolution as the actual data (but 
with much longer duration to improve the statistics).  
According to the binomial probabilities, we produced a 
piled-up light curve.  The 2 adjustable parameters are the
input count rate (before pile-up) and the probability that 
any pair of events occurring in the same time bin will pile-up.  
We adjusted these parameters to match the actual ACIS 
observation as shown in Figure~\ref{fig:pileup}.  We produced 
a Leahy normalized power spectrum using the simulated light 
curve, and the measured Poisson level is 1.78.

Simulation \#2:  In addition to the deadtime caused by pile-up, 
there is a fixed, instrumental deadtime that is independent of 
the event rate and is set by the CCD readout time of 41.04
milliseconds.  To determine the effect of the readout time, 
we simulated a light curve where there is a probability 
(equal to the readout time divided by the 0.4~s exposure
time per frame) that a photon that hits the detector will 
not be counted.  We produced a Leahy normalized power spectrum
using the simulated light curve, and the Poisson level is 1.97.

Simulation \#3:  Here, we made simulated light curves 
similar to those used for simulation \#1 except that we 
included white noise in excess of the Poisson noise.  
We produced a light curve free of pile-up and also a 
light curve with pile-up and made rms normalized power
spectra for each.  The fractional rms amplitude changes 
from 19\% to 18\% when pile-up effects are included.  
The shapes of the power spectra with and without pile-up
are not significantly different.

In summary, these simulations demonstrate that the low
Poisson noise level measured for the actual data is mainly
due to pile-up effects, and there is also a small drop 
due to the CCD readout time.  For noise levels similar
to those we measure for observation 3 ($\sim$18\%), pile-up
only causes the fractional rms amplitude to drop by about
1\%, and the pile-up does not change the shape of the
power spectrum.

\begin{center}
{\small APPENDIX B}
\end{center}

\begin{figure}
\centerline{\includegraphics[width=0.5\textwidth]{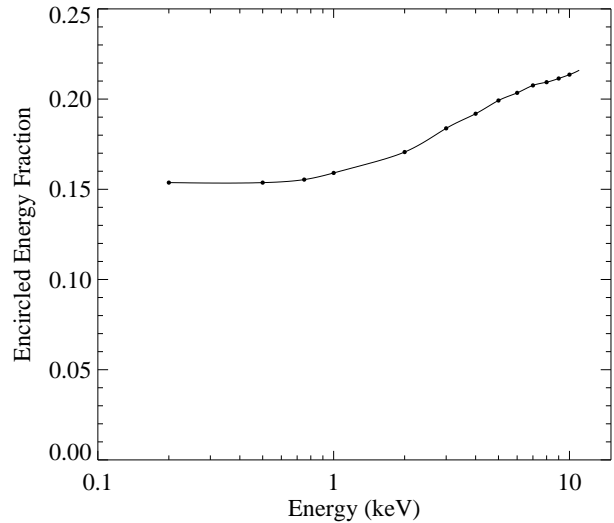}}
\vspace{0.2cm}
\caption{ChaRT ({\em Chandra} Ray Tracing) results for a 
source $2^{\prime}.7$ off-axis showing the fraction of 
counts in the elliptical annulus we used to extract the 
observation 3 ACIS spectrum (see text for the precise 
region parameters) as a function of energy.  The points
come from directly from the ChaRT simulations, and the
solid line is the spline interpolation that was applied 
to the ACIS response matrix for observation 3 to account 
for the energy dependence of the {\em Chandra} PSF.\label{fig:psf}}
\end{figure}

Here, we describe how we corrected the observation 3 ACIS 
response matrix for the energy dependence of the {\em Chandra} 
PSF.  We generated this correction using the web-based 
{\em Chandra} calibration tool ``ChaRT'' ({\em Chandra} 
Ray Tracing).  For 13 different energies from 0.2 to 10 keV, 
we used ChaRT to simulate a monochromatic source at the 
position of XTE J1650--500, $2^{\prime}.7$ off-axis.  For
each energy, we determined the fraction of counts in the
same elliptical annulus that we used to extract the 
XTE J1650--500 spectrum (see text), and the fractions are 
shown in Figure~\ref{fig:psf}.  We then used a spline 
interpolation to determine the fraction for each energy 
bin in the response matrix, and we applied the interpolated
values to the response matrix.  While this correction is 
necessary, we note that, in this case, it only changes the 
measurement of the power-law index from $\Gamma = 1.79\pm 0.13$ 
to $\Gamma = 1.93\pm 0.13$.  Although we do not have a method 
of accurately determining the systematic error on this correction, 
even if the correction is only good to 10-20\%, the systematic 
error is much smaller than the statistical error.

\clearpage

\begin{table}
\caption{{\em Chandra} and {\em RXTE} Observations\label{tab:obs}}
\begin{minipage}{\linewidth}
\footnotesize
\begin{tabular}{c|c|c|c|c|c} \hline \hline
Obs./   & & & Energy Band & Exposure & Count\\
UT Date & MJD Start\footnote{Modified Julian Date (JD-2400000.5) at exposure midpoint.} & 
Instrument & (keV) & (s) & Rate\footnote{The MEG and HEG rates quoted are for the +1 and
$-1$ grating orders combined.}\\ \hline
1                 & 52298.00688 & ACIS/HETG/MEG  & 0.7-7 & 10001 & $0.479\pm 0.007$\\
(2002 Jan. 23-24) & 52298.00688 & ACIS/HETG/HEG  & 1-9   & 10001 & $0.271\pm 0.005$\\
                  & 52298.03298 & {\em RXTE}/PCA & 3-20  &  1904 & $9.42\pm 0.14$\\ \hline
2                 & 52309.62048 & ACIS/HETG/MEG  & 0.7-7 &  9509 & $0.457\pm 0.007$\\
(2002 Feb. 4)     & 52309.62048 & ACIS/HETG/HEG  & 1-9   &  9509 & $0.248\pm 0.005$\\
                  & 52309.71204 & {\em RXTE}/PCA & 3-20  &  1872 & $8.69\pm 0.14$\\ \hline
3                 & 52335.10444 & ACIS\footnote{For this observation, we used a 1/8 
subarray and an offset pointing of $2^{\prime}.7$ to mitigate the effects of photon 
pile-up.}         & 0.5-8 & 18266 & $0.493\pm 0.005$\footnote{This is the count rate 
for the source, including the core, which is moderately affected by photon pile-up.  
We estimate that the count rate would be $\sim$15\% higher without pile-up.}\\
(2002 Mar. 2)     & 52335.15312 & {\em RXTE}/PCA & 3-20  &  2384 & $1.38\pm 0.11$\\ \hline
\end{tabular}
\end{minipage}
\end{table}

\begin{table}
\caption{Parameters from {\em Chandra} and PCA Spectral Fits\label{tab:spec}}
\begin{minipage}{\linewidth}
\footnotesize
\begin{tabular}{c|c|c|c|c|c} \hline \hline
 & $N_{\rm H}$ & & Flux & & Fe K$\alpha$ Line EW\\
Obs.\footnote{These are parameters from absorbed power-law fits to the spectra.
All errors and upper limits are 90\% confidence ($\Delta\chi^{2} = 2.7$).} & 
($10^{21}$ cm$^{-2}$) & $\Gamma$ & (1-9 keV)\footnote{The 1-9 keV absorbed flux 
in units of erg cm$^{-2}$ s$^{-1}$.} & $\chi^{2}/\nu$ & Upper Limit (eV)\footnote{90\%
confidence upper limit on the equivalent width of a narrow emission line at 6.4 keV.}\\ \hline
1 & $6.7\pm 0.5$ & $1.66\pm 0.05$ & $4.76\times 10^{-11}$ & 116/155 & $< 162$\\
2 & $6.7\pm 0.5$ & $1.66\pm 0.05$ & $4.46\times 10^{-11}$ & 164/155 & $< 169$\\
3 & $6.4\pm 0.7$ & $1.93\pm 0.13$ & $7.9\times 10^{-12}$ & 104/100 & $< 212$\\ 
3 & 6.7\footnote{Fixed.} & $1.96\pm 0.09$ & $8.0\times 10^{-12}$ & 104/101 & $< 212$\\
\hline
\end{tabular}
\end{minipage}
\end{table}

\begin{table}
\caption{Paramaters for the Power Spectra\label{tab:power}}
\begin{minipage}{\linewidth}
\footnotesize
\begin{tabular}{c|c|c|c|c|c} \hline \hline
Obs.\footnote{These are parameters from fits with a zero-centered Lorentzian.
Errors are 68\% confidence.} & Instrument & Energy Band (keV) & FWHM/2\footnote{The half-width of the zero-centered 
Lorentzian in Hz.} & Fractional RMS (\%) & $\chi^{2}/\nu$\\ \hline
1+2\footnote{A QPO may be present near 0.11 Hz with a fractional rms of about
12\%.  The significance of the QPO is only required at a confidence level of
between 2-$\sigma$ and 3-$\sigma$.} & PCA & 3-20 & $0.067\pm 0.007$ & $33.1\pm 1.1$ & 93/72\\
1+2 & MEG+HEG & 0.7-9 & $0.079\pm 0.020$ & $30.1\pm 2.5$ & 25/23\\
3 & ACIS & 0.5-8 & $0.0035\pm 0.0010$ & $19^{+7}_{-3}$ & 56/65\\ \hline
\end{tabular}
\end{minipage}
\end{table}

\end{document}